\documentclass[12pt]{article}
\usepackage[margin=1.25in]{geometry}
\usepackage{amsmath,amsthm,amssymb}
\usepackage{graphicx}
\usepackage{wasysym}
\usepackage{todonotes}
\usepackage[linesnumbered,ruled,vlined]{algorithm2e}
\usepackage{setspace}
\usepackage{url}
\usepackage{hyperref}

\newcommand{\B}{\mathbb}

\newcommand{\R}{\B{R}}

\global\long\def\state{\operatorname{state}}
\global\long\def\conv{\operatorname{conv}}

{\bfseries}{\rmfamily}
\newtheorem{definition}{Definition}[section]
\newtheorem{theorem}{Theorem}[section]
\newtheorem{proposition}[theorem]{Proposition}

\newtheorem{lemma}[theorem]{Lemma}

\usepackage{xcolor}
\definecolor{light-gray}{gray}{0.95}

\usepackage{subfiles}

\title{Expiring Assets in Automated Market Makers}
\author{
Kenan Wood\\
\texttt{kewood@davidson.edu}\\
Davidson College
\and
Maurice Herlihy\\
\texttt{mph@cs.brown.edu}\\
Brown University
\and
Hammurabi Mendes\\
\texttt{hamendes@davidson.edu}\\
Davidson College
\and
Jonad Pulaj\\
\texttt{jopulaj@davidson.edu}\\
Davidson College
}
\date{January 2024}

\begin{document}

\maketitle


\begin{abstract}
  An automated market maker (AMM) is a state machine that manages pools of assets, allowing parties to buy and sell those assets according to a fixed mathematical formula.
  AMMs are typically implemented as smart contracts on blockchains, and its
  prices are kept in line with the overall market price by \emph{arbitrage}:
  if the AMM undervalues an asset with respect to the market,
  an ``arbitrageur'' can make a risk-free profit by buying just enough of that asset
  to bring the AMM's price back in line with the market.
    
  AMMs, however, are not designed for assets that \emph{expire}:
  that is, assets that cannot be produced or resold after a specified date.
  As assets approach expiration,
  arbitrage may not be able to reconcile supply and demand,
  and the liquidity providers that funded the AMM
  may have excessive exposure to risk due to rapid price variations.

  This paper formally describes the design of
  a decentralized exchange (DEX) for assets that expire,
  combining aspects of AMMs and limit-order books.
  We ensure \emph{liveness} and \emph{market clearance},
  providing mechanisms for liquidity providers to
  control their exposure to risk and
  adjust prices dynamically in response to situations where arbitrage may fail.
\end{abstract}

\section{Introduction}
An \emph{automated market maker} (AMM)~\cite{BartolettiCL2021} is an automaton
(usually implemented as a smart contract)
that trades one electronic asset for another at rates
set by a fixed mathematical formula.
\emph{Liquidity providers} (LPs) lend assets to the AMM
in return for a share of transaction fees and governance rights.
Today, AMMs form a multi-billion dollar business~\cite{xu2023sok}.
The principal benefit of AMMs over more traditional order-book
methods is that an AMM offers its assets at a take-it-or-leave-it price,
so trades occur immediately,
without need for complex bidding strategies.

AMMs typically trade long-lived assets such as cryptocurrencies
or stablecoins.
In this paper, we investigate a decentralized exchange (DEX) that can be
designed to trade \emph{expiring assets},
those that have a fixed date after which they cannot be resold.
For example,
a flight ticket from New York to Paris
that departs on Monday cannot be sold on Tuesday.
Expiration can have complicated effects on price:
a last-minute airline ticket may sell at a premium,
but a last-minute ticket for a Broadway show may sell at a discount.

Parties can play three distinct roles.
\emph{Producers} sell (expiring) \emph{tokens} through the DEX.
For ease of exposition,
all tokens expire at the same well-known time and date.
\emph{Consumers} use (electronic) \emph{cash} to purchase those tokens through the DEX.
Cash has a stable value, and acts as the num\'eraire.
(Some parties may combine both roles to act as \emph{arbitrageurs},
buying and selling tokens with the effect of aligning
the DEX token price with an exogenous market price.)
Finally, \emph{liquidity providers} (LPs) are market makers,
buying tokens from producers and selling them to consumers
in return for fees.

For most of the DEX's lifetime,
it acts like a standard automated market maker.
Producers and consumers sell and buy tokens at a take-it-or-leave-it
price determined by the formula specific to the DEX.
All transactions are executed immediately.
If there is a known, exogenous ``market'' where tokens can also be bought and sold,
then arbitrageurs keep the DEX price in agreement with the market price.
Liquidity providers bootstrap the DEX by loaning assets,
and they profit by taking a fee for each transaction.
When the the system shuts down, its remaining cash is returned to the LPs.

As the expiration date approaches,
standard market mechanisms may fail.
For some unknown but brief time before token expiration,
token supply may vanish:
producers stop selling new tokens on the DEX, no matter the price.
(For example, airlines cannot create more seats for an imminent Monday flight to Paris.)
Similarly, token demand may vanish:
consumers stop buying tokens on the DEX, no matter the price.
(Hotel rooms in the Caribbean are not going to be booked by people that cannot get there on time)
%
In addition, airline tickets, hotel rooms, and similar expiring assets have prices that can vary highly in response to exogenous events that may be harder to predict, such as weather, geopolitics, etc.
Under these circumstances,
the DEX can no longer rely on arbitrage to keep the market efficient.

In particular, conventional AMMs fail under those circumstances.
Normally,
if an AMM's price for tokens (in cash) rises above the market price,
then an arbitrageur would buy (or create) new tokens
and sell them to the AMM, thus collecting a risk-free profit
and bringing the AMM's token price back to the market rate.
If supply vanishes, however,
each token purchase will drive the AMM price up,
and eventually the AMM's token price will rise just above the market price,
and the remaining tokens will expire unsold,
even though both the consumers and the LPs would have been better off selling those tokens at a lower price.
From this we learn that
a DEX for expiring assets cannot rely exclusively on arbitrage
to keep prices in line with the market.

In an alternative scenario, if demand vanishes,
token producers will dump their expiring tokens on the AMM.
Of course, such panicky selling will cause the AMM price to fall,
but the token producers will realize at least some cash for their tokens.
But these trades come at the expense of the LPs who loaned the
cash that is being replaced with soon-to-be worthless tokens.
From this we learn that
AMMs for expiring assets must provide some way for LPs to intervene to limit
their exposure to market distortions
due to expiration.

This paper describes the design of a decentralized exchange (DEX)
for tokens that expire, with the following design goals.

\begin{itemize}
\item
  \emph{Market clearing}: at expiration time,
  it is never the case
  that there is a producer willing to sell a token at some price
  \footnote{
A producer may have a minimum price even for expiring tokens.
For example, a hotel might not rent a room for less than the cost
of cleaning that room.},
  and a consumer willing to buy at that price,
  yet that token remains unsold,
  
\item
  \emph{Instant gratification}:
  each asset is bought and sold either immediately
(during normal operations)
or after a known delay (as expiration approaches), and

\item
  \emph{Incentives}:
  each participant has an incentive to participate.
\end{itemize}





This paper's contribution is the design and analysis
of a novel DEX for trading assets that expire.
Our analysis points out ways in which conventional AMM design
fails to address issues raised by expiring tokens,
and we propose the following mechanisms to augment
conventional AMMs to handle expiring assets.
\begin{itemize}
\item
  LPs can respond to unforeseen fluctuations in token supply and demand -- in particular, in cases resulting in arbitrage failure --
  by directly intervening to change the shape of the AMM's curve.
  
\item
  LPs can limit its exposure
  to demand failure by setting a token price below which
  their cash cannot be used to buy tokens
  (similar to Uniswap V3's concentrated liquidity).
\item
  The AMM is augmented with a pair of order books to ensure market clearing, even in situations of supply and demand failure. For example, just before the tokens expire,
  any tokens remaining in the DEX are auctioned off, prioritizing LPs in order of cumulative participation.
\end{itemize}


LPs are assumed to be rational.
They may disagree on market predictions, but they do not act maliciously against one another.

\section{Related Work}
Today, the most popular automated market maker is
\emph{Uniswap}~\cite{uniswapv2,AngerisKCNC2019,uniswapv3,zhang2018},
a family of constant-product AMMs.
Originally trading between ERC-20 tokens and ether cryptocurrency,
later versions added direct trading between pairs of ERC-20 tokens,
and allowed liquidity providers
to restrict the range of prices in which their assets participate.
\emph{Bancor}~\cite{bancor} AMMs permit more flexible pricing schemes,
and later versions~\cite{bancorv2} include integration with external
``price oracles'' to keep prices in line with market conditions.
\emph{Balancer}~\cite{balancer} AMMs trade across more than two assets,
based on a \emph{constant mean} formula that generalizes constant product.
\emph{Curve}~\cite{curve} uses a custom curve specialized for trading
\emph{stablecoins}, maintaining low slippage and divergence loss
as long as the stablecoins trade at near-parity.
The formal model for AMMs used here
is adapted from Engel and Herlihy~\cite{EngelH2021}.

Xu \emph{et al.}~\cite{xu2023sok} and 
Bartoletti \emph{et al.}~\cite{BartolettiCL2021}
provide informative overviews on AMM protocols.
Angeris and Chitra~\cite{AngerisC2020}
introduce a \emph{constant function market maker} model
and focus on conditions that ensure that agents who
interact with AMMs correctly report asset prices.

Aoyagi~\cite{Aoyagi2020}
analyzes strategies for constant-product AMM liquidity providers
in the presence of ``noise'' trading, which is not intended to move prices,
and ``informed'' trading, intended to move the AMM to the stable point
for a new and more accurate valuation.
Angeris \emph{et al.}~\cite{AngerisEC2020}
propose an economic model relating how the curvature of the AMM's
function affects LP profitability in the presence of
informed and noise traders.
Bichuch and Feinstein~\cite{bichuch2023axioms} propose a general mathematical framework for AMMs,
and Capponi \emph{et al.}~\cite{capponi2021adoption} analyze AMMs using a game theoretic model.

Our mechanism where liquidity providers can \emph{freeze} their liquidity under adverse market conditions is similar to Uniswap V3's~\cite{uniswapv3} notion of \emph{concentrated liquidity},
though the purpose and operational details are somewhat different.

Ramseyer \emph{et al.}~\cite{RamseyerGGM2023}
describe several ways to integrate constant-function AMMs with batch auctions.
Like the exchange described in this paper,
their exchanges have hybrid structures,
but their goals and techniques are substantially different,
as they are concerned with broad economic properties,
not with expiring assets.

\section{Informal System Overview}
\label{Sec-Overview}
Our DEX is composed of an AMM and two order books that operate under specific rules. We give now an operational overview of our DEX, starting with the AMM component. We point the reader to the appropriate sections containing detailed discussion as we present the concepts below.

An AMM owns and trades two kinds of assets: a stable asset $X$
(informally called \emph{cash}),
and an expiring asset $Y$ (informally called \emph{tokens}),
These assets are loaned to the AMM by \emph{liquidity providers} (LPs).
As explained later,
LPs can be \emph{active} or \emph{inactive}.

The AMM tracks a pair $(x,y) \in \R_{> 0}^2$,
where $x$ (respectively $y$) is the amount of $X$ (respectively $Y$)
that the AMM offers for trading.

Any party $\ell$ can become a liquidity provider by lending assets to the AMM.
In return, $\ell$ receives a \emph{share}
$s_\ell \in (0, 1]$
proportional to its contribution.
In departure from ``pure'' AMM designs,
an LP can dynamically influence AMM prices to a degree
proportional to its share. 
The particular protocol and curve design is discussed in Section~\ref{Sec-PriceAdjustment}.

LPs can withdraw liquidity at well-defined times,
called \emph{epoch boundaries}.
LPs can make withdrawals denominated in both types of assets,
or in only the stable asset $X$.
LPs can \emph{freeze} their liquidity,
in a manner comparable to Uniswap V3's
notion of concentrated liquidity~\cite{uniswapv3}.
An LP $\ell$ can set a minimum spot price for $Y$ tokens in terms of $X$ cash.
If the AMM's spot price for tokens falls below that LP's minimum,
that LP becomes \emph{inactive}
and its liquidity is not used in trades.
Of course, LPs do not accrue fees while they are inactive.
An LP can use freezing to protect itself if it fears that demand has collapsed,
causing producers to flood the AMM with $Y$ tokens no one will buy.
If the price later rises above an LP's minimum, that LP becomes active again. Details are discussed in Section \ref{Sec-FreezingAndRelated}.

To ensure market clearing,
there are complementary buyer and seller order books
that are primarily used in the following situations: 
just before the tokens expire, 
when there are not enough tokens in the system to execute any more cash-for-tokens trades (demand is too high), 
or when all LPs are inactive, so no more AMM trades can be executed temporarily.
Details can be found in Section~\ref{Sec-MarketClearing}.
If a consumer registers a below-spot-market bid (on $Y$ tokens)
on the buyer order book and is still present near the expiration time,
that bid is executed with a second price auction.
There is also a seller order book,
(almost) symmetric to the buyer order book,
where sellers offer to sell $Y$ tokens that remain after the AMM token pool is exhausted.
The combination of the AMM and the two order books acts like an AMM while normal supply-and-demand laws hold,
but acts more like an order book (or auction) near the expiration date when those laws may fail.

We also include a mechanism to incentivize LP participation, so that the AMM (instant-gratification) component of the DEX remains live for a longer period of time. Of course, LPs can withdraw or freeze liquidity. However, if an LP $\ell$ chooses to take on more risk by participating more, they will be given priority in clearing their own tokens, and with a higher price, in the described market clearing auction. Thus, even if two LPs $\ell_1, \ell_2$ have the same share and hold the same number of tokens, if $\ell_1$ participates in many more (and larger) trades than $\ell_2$, then $\ell_1$ will receive a greater share of the auction proceeds than $\ell_2$. We discuss these details in Section~\ref{Sec-MarketClearing}.

\subsection{Classifying Consumers}
In standard AMM models (\emph{e.g.}, Milionis \emph{et al.}~\cite{MilionisMRZ2022}),
so-called \emph{noise} traders buy tokens with the intent of consuming
them when they expire,
while \emph{arbitrageurs} (sometimes called \emph{informed} traders)
trade between the DEX and a centralized market to make a profit.
These distinctions still apply when assets expire;
but consumers, producers, and even LPs have important additional characteristics.
\begin{itemize}
\item
  \emph{Price}:
  How much are customers willing to pay for tokens
  as expiration approaches?
  If demand outlasts supply,
  then producers may be unable to bring new tokens to market,
  driving up the price as time runs out.
  If supply outlasts demand,
  producers may flood the DEX with unwanted tokens,
  yielding a final price of (practically) zero. The design of the DEX should be able to accommodate both extremes as well as outcomes in between.
\item
  \emph{Urgency}:
  What kinds of risks influence consumer behavior?
  A consumer who waits until the last minute
  may not be able to buy a token if demand outstrips supply
  (\emph{e.g.,} all flights to Paris on Monday are sold out).
  Alternatively, a last-minute consumer may receive a favorable deal if supply outstrips demand
  (\emph{e.g.,} a last-minute bargain on a theater ticket).
  Consumers worried about missing out are said to have \emph{high urgency},
  while those willing to gamble on last-minute bargains have \emph{low urgency}.
\end{itemize}
To summarize, there are three kinds of consumers:
\begin{itemize}
\item
  \emph{Bargain hunters}: low urgency, low price.
  For example, a consumer who is willing to buy a last-minute ticket to a Broadway show if those tickets are cheap enough.
\item
  \emph{Normal customers}: high urgency, low price.
  These are conventional AMM (noise or informed) traders who want to execute their trades immediately at a take-it-or-leave it price.
\item
  \emph{High flyers}: high urgency, high price.
  For example, a consumer who suddenly needs to fly to an urgent business meeting in Paris on Monday. 
\end{itemize}
We do not consider consumers with low urgency but high price.
A symmetric classification can be applied to producers,
but we focus mostly on consumers.

The last-minute consumer order book allows bargain hunters
to bid on tokens at prices below the AMM's asking price.
A bargain hunter unwilling to meet the AMM's asking price
can register a bid on the last-minute consumer order book.
That bid will be executed immediately if
the AMM price falls to the bid price.
Just before expiration,
any unsold tokens will be distributed among the order book bidders
via a second-price auction
(or any other incentive-compatible auction mechanism).

Normal consumers,
who want to buy tokens at market price right away,
and normal producers,
who want to sell tokens at market price right away
both use the AMM component of the DEX.

High flyers,
who need tokens at (almost) any price,
will first go to the AMM.
If the AMM is sold out of tokens
(because, for example, expiration is near
and supply is exhausted),
then a high-flyer can place an order on the last-minute consumer order book.
Last-minute producers who consider the AMM price too low
can monitor this order book and fulfill any satisfactory orders.

The last-minute producer and consumer order books
ensure \emph{market clearing}:
just before expiration,
no token remains unsold if its producer was willing to sell
at a price some unfulfilled consumer was willing to pay.

As discussed below,
additional mechanisms are necessary to serve the interests of LPs,
who are also exposed to risks stemming from failures in either supply or demand as expiration approaches.

\section{Mathematical Preliminaries}
We now prepare to describe the system in proper detail, first quickly discussing preliminary notation and concepts that are necessary for a formal system presentation in the sections that follow.

For vectors $x = (x_1, \dots, x_n) \in \R^n$ and $y = (y_1, \dots, y_n) \in \R^n$, we write $x \le y$ provided $x_i \le y_i$ for all $i \in [n]$, where $[n] = \{1, \dots, n\}$; if also $x \ne y$, we write $x \lneq y$.
When referencing topological properties in this paper, we assume that any subset of $\R^n$ (for any $n \ge 1$) is given the subspace topology, unless otherwise specified.
We define $\R_{> 0} = \{x \in \R: x> 0\}$. A function $A: \R^n \to \R$ is \emph{twice-differentiable} if all of its second partial derivatives exist and are continuous.
We also say that $A$ is \emph{strictly increasing in each coordinate} if for all $x,y \in \R^n$ such that $x \lneq y$, we have $A(x) < A(y)$.

A set $\Omega \subseteq \R^n$ is \emph{convex} if for all distinct $x,y \in \Omega$ and $t \in (0, 1)$, we have $tx + (1-t)y \in \Omega$; if every such vector $tx + (1-t)y$ is in the interior of $\Omega$, we say $\Omega$ is \emph{strictly convex}. It follows that that any convex open set is strictly convex. Given a set $S \subseteq \R^n$, the \emph{convex hull} of $S$, denoted $\conv S$, is defined as the intersection of all nonempty convex sets $\Omega \subseteq \R^n$ that contain $S$.
Suppose $\Omega \subseteq \R^n$ is convex. A function $f: \Omega \to \R$ is said to be \emph{strictly convex} provided that for every distinct $x,y \in \Omega$ and $t \in (0,1)$, it follows that $f(tx + (1-t)y) < tf(x) + (1-t)f(y)$.


\section{Price Adjustment Mechanisms}
\label{Sec-PriceAdjustment}
As discussed in Section~\ref{Sec-Overview}, our DEX has an AMM meant to provide an instant buying price for consumers. 
In a conventional constant-function AMM, token prices are kept consistent with the market by external \emph{arbitrageurs}; if the token price rises above the market price, an arbitrageur may sell overpriced tokens to the AMM, making a risk-free profit. Conversely, if the AMM price falls below the market price, an arbitrageur may buy the bargain tokens from the AMM and resell them on the market. Arbitrage works well for a conventional AMM because there are sufficiently many tokens available to be traded, so that arbitrageurs are always able to trade in the direction needed to bring the AMM price back in line with the market. 
But close to expiration date, token supply or demand may vanish, and the normal price regulation cannot take place. For example, airlines will not schedule new seats to Paris as the flight time becomes too close. Similarly it makes little sense to book a room in the Caribbean starting in one hour if the room is five hours away. In addition, airline tickets, hotel rooms, event seats, and similar expiring assets have prices that may be influenced by hard-to-predict, exogenous events, such as weather, geopolitical events, etc.

In conventional AMMs,
LPs can deposit or withdraw liquidity,
but are otherwise passive,
allowing producers and consumers to set asset prices via trading.
As noted, however,
arbitrageurs may become unable
to intervene to keep prices close to market levels as expiration approaches.
So we propose a mechanism to allow LPs to intervene actively
to adjust the curve when conventional methods are in danger of failing.

More specifically, in the period while expiration is far off,
when the DEX acts like a conventional AMM,
arbitrageur profits come at the expense of LPs,
a cost known as \emph{divergence loss}.
The LPs effectively gamble that asset prices will be stable enough
that the fees collected will outweigh any divergence loss.
As expiration approaches, however, prices may become volatile and
arbitrage may become ineffective,
so LPs require the ability protect their investments by intervening directly to adjust prices (also to actively protect their capital -- see Section~\ref{Sec-FreezingAndRelated}),
effectively assuming the role of conventional arbitrageurs,
but without divergence loss.

The duration between when the DEX is started and when the token expires is divided into disjoint intervals called \emph{epochs},
separated by \emph{epoch boundaries}.
There are two special epoch boundaries:
$t_e$, when the tokens expire,
and $t_r < t_e$, when the AMM halts and the last-minute auctions occur. Let $T$ be the set of all epoch boundaries strictly before $t_r$.
 

We say that a \emph{state curve} is a strictly decreasing homeomorphism $f: \R_{> 0} \to \R_{> 0}$. If $x > 0$, we say that the pair $(x, f(x))$ is a \emph{state} of $f$, and we denote the set of all states of $f$ by $\state(f)$. Operationally, we say that the AMM is in state $(x,y) \in \state(f)$ if it contains $x$ units of $X$ and $y$ units of $Y$. Trades change the amounts $x$ and $y$ of $X$ and $Y$, respectively, so that $(x,y)$ is always in $\state(f)$.

We consider only a fixed subfamily of possible state curves. Let $0 < a < b$ be fixed positive lower and upper bounds. (We may wish to require that $a < 1$ and $b = \frac{1}{a}$ for symmetry, though this is not required.) For every $c \in [a,b]$ and possible state $(x_0,y_0) \in \mathbb{R}_{> 0}^2$, define the state curve
$$f_{c,x_0,y_0}(x) = y_0 \left(\frac{x}{x_0} \right)^{-c}$$
for all $x > 0$. Then $(x_0, y_0) \in \state(f_{c, x_0,y_0})$. Let $L$ be the current set of \emph{active} liquidity providers (whose assets are allocated to trading), and $L'$ is the set of $\emph{inactive}$ liquidity providers (whose assets are frozen). The sets $L$ and $L'$ and their associated methods are formally defined in Section \ref{Sec-FreezingAndRelated}. 

At each time in $T$, each liquidity provider $\ell \in L \cup L'$ chooses some $c_\ell \in [a,b]$ that informally represents the price $\ell$ believes is most economically efficient, with knowledge of the current state of the AMM. Each $c_\ell \in [a,b]$ is aggregated into a single $c \in [a,b]$ with some deterministic \emph{aggregation algorithm} $\mathtt{aggregate}()$ by computing $c \gets \mathtt{aggregate}()$ at each epoch boundary. The $\mathtt{aggregate}()$ algorithm has parameters $(c_\ell)_{\ell \in L}$ and $(s_\ell)_{\ell \in L}$\footnote{We exclude the $c_\ell$'s and shares of inactive liquidity providers because their share of liquidity is not well-defined, as it is not necessarily the same over both of the assets. See Section \ref{Sec-FreezingAndRelated} for details.} and returns some element of $[a,b]$. For fixed $L$, since $\mathtt{aggregate}()$ is deterministic, it can be viewed as a map 
\[
\mathtt{aggregate}: [a,b]^{L} \times \left\{(s_\ell)_{\ell \in L} \in [0,1]^{L}: \sum_{\ell \in L} s_\ell = 1\right\} \to [a,b],
\]
where $A^B$ is the set of all tuples indexed by $B$ with values in $A$ for any sets $A,B$. Now we define the soundness properties for an aggregation algorithm.
\begin{definition}\label{aggregation}
    Let $\mathtt{aggregate}()$ be an aggregation algorithm; let $L$ be the particular set of active LPs at some time, and consider the corresponding aggregation function. We say that $\mathtt{aggregate}()$ is \emph{valid} if the following axioms hold.
\begin{enumerate}
    \item For each $\ell \in L$, the function $\mathtt{aggregate}$ has continuous partial derivatives with respect to $c_\ell$. Additionally, 
    \[
    \frac{\partial (\mathtt{aggregate})}{\partial\log(c_\ell)} = s_\ell \cdot \mathtt{aggregate}((c_\ell)_{\ell \in L}, (s_\ell)_{\ell \in L}).
    \]
    \item 
    If there exists some $x$ such that $c_\ell=x$ for all $\ell \in L$, then 
    \[
    \mathtt{aggregate}((c_\ell)_{\ell \in L}, (s_\ell)_{\ell \in L}) = x.
    \]
\end{enumerate}
\end{definition}

For brevity, let $c = \mathtt{aggregate}((c_\ell)_{\ell \in L}, (s_\ell)_{\ell \in L})$.

The first condition implies that for a small fixed change in $\log c_\ell$ from an LP $\ell$ with share $s_\ell$, the corresponding change in $c$ is proportional to $s_\ell$. We use a logarithmic scale since derivatives are limits of \emph{additive changes}, and $\log$ is an isomorphism from the group $\R_{>0}$ under multiplication to the group $\R$ under addition. (Note that $c_\ell \in \R_{>0}$, so it only makes sense to give them a multiplicative structure.)

The second condition is a simple boundary condition to guarantee that if all LPs are in consensus with a particular value $c$, the aggregated constant is the common value $c$.

In our model, we define $\mathtt{aggregate}((c_\ell)_{\ell \in L}, (s_\ell)_{\ell \in L})$ to return
\begin{equation}
\label{eqn:c}
    \prod_{\ell \in L} c_{\ell}^{s_{\ell}},
\end{equation}
the geometric mean of each $c_\ell$ weighted by the share $s_\ell \in [0,1]$ each liquidity provider holds in the system. Interestingly, we show in the Appendix that this aggregation algorithm is the \emph{unique} valid aggregation algorithm.

The state curve is implied from the global variables $c, x_0, y_0$ stored in the AMM. Until the next time in $T$ in which liquidity providers may update their value of $c_\ell$, we fix $c$ as defined above.
Note that when the state $(x,y)$ changes from adding/removing liquidity, freezing/unfreezing liquidity, and liquidating (specified in Section \ref{Sec-FreezingAndRelated}), we update $x_0 \gets x$ and $y_0 \gets y$, and we must also update the state curve in these cases. This is necessary because the liquidity state $(x,y)$ after completing any of these methods is no longer in the state space of the current state curve $f_{c, x_0, y_0}$. We omit updates to $(x_0, y_0)$ in any pseudocode of these methods for simplicity of presentation.

\subsection{State Curve Properties}
Now, let us show that this state curve construction acts like an AMM. We use axioms proposed by Engel and Herlihy \cite{EngelH2021} that define an \emph{AMM} as follows. We note that this definition is only used in this section, and other references to the AMM refer to the system model described in this paper.

\begin{definition}\label{AMM_def}
    An \emph{AMM} is a function $A: \mathbb{R}_{>0}^2 \to \mathbb{R}$ that satisfies the following:
    \begin{itemize}
        \item $A$ is twice-differentiable;
        \item $A$ is strictly increasing in each coordinate;
        \item For each $b \geq 0$, the set $\{(x, y) \in \mathbb{R}_{>0}^2: A(x,y) \geq b\}$ is closed and strictly convex.
    \end{itemize}
\end{definition}
Under this definition, the state space of $A$ is the set $\{(x,y) \in \mathbb{R}_{>0}^2: A(x,y) = 0\}$ and is denoted $\state(A)$. This is the same notation as the state space of a state curve, but the definition of state space should be clear, according to whether the function is a state curve or an AMM. Henceforth, we use $A$ to denote an AMM and $f$ to denote a state curve. It is important to note the difference between an AMM and a state curve as defined previously; an AMM is a function $\R_{>0}^2 \to \R$, but a state curve is a function $\R_{>0} \to \R_{>0}$. An AMM captures more information than the set of states alone on a state curve alone. In particular, if $A$ is an AMM, then we may construct its state curve by mapping $x > 0$ to the unique $y > 0$ such that $A(x, y) = 0$. However, the converse is not necessarily true as not every state curve can be extended to an AMM satisfying the above three properties. Observe that there is no operational difference between an AMM and a state curve since the set of states completely determines the behavior of the system. This motivates the following definition.

\begin{definition}
Given state curve $f$, we say that $f$ \emph{induces an AMM} if there exists an AMM $A$ such that $\state(f) = \state(A)$.
\end{definition}

We will prove that for any $(x_0, y_0) \in \R_{> 0}^2$ and $c > 0$ that $f_{c, x_0, y_0}$ induces an AMM. For the remainder of this section, fix $(x_0, y_0) \in \R_{> 0}^2$ and $c > 0$. To prove that $f_{c, x_0, y_0}$ induces an AMM, we let $A(x, y) = x^cy - x_0^cy_0$. We shall first prove that $A$ is an AMM. It is easy to see the following two lemmas, and the third one is simple.
\begin{lemma}\label{differentiable}
    $A$ is twice differentiable.
\end{lemma}
\begin{proof}
    Observe that $\frac{\partial A}{\partial x} = cx^{c-1}y$ and $\frac{\partial A}{\partial y} = x^c$. Then $\frac{\partial^2 A}{\partial x^2} = c(c-1)x^{c-2}y$ and $\frac{\partial^2 A}{\partial y^2} = 0$, which are both continuous. Since both second partial derivatives of $A$ exist and are continuous, $A$ is twice differentiable.
\end{proof}
\begin{lemma}\label{increasing}
    $A$ is strictly increasing in each coordinate.
\end{lemma}
\begin{proof}
    If $x' > x > 0$ and $y > 0$, then since $c > 0$, $A(x', y) = (x')^cy-x_0^cy_0 > x^cy - x_0^cy_0 = A(x,y)$. Similarly, if $x > 0$ and $y' > y > 0$, then $A(x,y') = x^cy' - x_0^cy_0 > x^cy - x_0^c y_0 = A(x,y)$, as desired.
\end{proof}
\begin{lemma}\label{convex}
    For any $b \ge 0$, the set $\{(x,y) \in \R_{>0}^2: A(x,y) \ge b\}$ is closed and strictly convex.
\end{lemma}
\begin{proof}
    Let $b \ge 0$ and let $S = \{(x,y) \in \R_{>0}^2: A(x,y) \ge b\}$. Since $A$ is continuous and $[b, \infty)$ is closed, $S = A^{-1}([b, \infty))$ is closed because the preimage of a closed set under a continuous function is closed. To prove that $S$ is strictly convex, it suffices to show that for any distinct $(x_1, y_1), (x_2, y_2) \in S$ and $t \in (0, 1)$, it follows that $A(t\cdot (x_1, y_1) + (1-t) \cdot (x_2, y_2)) > b$. To this end, notice that since $(x_1, y_1), (x_2, y_2) \in S$, we know $x_1^c y_1 \ge b+x_0^c y_0$ and $x_2^c y_2 \ge b+x_0^c y_0$. Thus, by the Weighted AM-GM Inequality \cite[p.~74]{cvetkovski2012inequalities},
    \begin{align*}
        A(t\cdot (x_1, y_1) + (1-t) \cdot (x_2, y_2)) &= (tx_1 + (1-t)x_2)^c (ty_1 + (1-t)y_2) - x_0^c y_0\\
        &> (x_1^t x_2^{1-t})^c (y_1^t y_2^{1-t}) - x_0^c y_0\\
        &= (x_1^c y_1)^t (x_2^c y_2)^{1-t} - x_0^c y_0\\
        &\ge (b + x_0^c y_0)^t (b + x_0^c y_0)^{1-t} - x_0^c y_0\\
        &= b.
    \end{align*}
    This shows that $S$ is a strictly convex set, as desired.
\end{proof}

Finally, we must verify that $f_{c, x_0, y_0}$ and $A$ have the same state spaces.

\begin{lemma}\label{curve_is_AMM}
    $\state(f_{c, x_0, y_0}) = \state(A)$.
\end{lemma}
\begin{proof}
    Observe that 
    \[
    (x, y) \in \state(f_{c, x_0, y_0}) \Leftrightarrow y = y_0(x/x_0)^{-c} \Leftrightarrow x^cy = x_0^c y_0 \Leftrightarrow (x,y) \in \state(A).
    \]
    Thus $\state(f_{c, x_0, y_0}) = \state(A)$.
\end{proof}

As a consequence, we have the following.
\begin{theorem}
The state curve $f_{c, x_0, y_0}$ induces an AMM.
\end{theorem}
\begin{proof}
    In letting $A(x, y) = x^cy - x_0^cy_0$ as above, Lemmas \ref{differentiable}, \ref{increasing}, and \ref{convex} imply that $A$ is an AMM. By Lemma \ref{curve_is_AMM}, it follows that $f_{c, x_0, y_0}$ induces an AMM.
\end{proof}
Since every state curve used in our price control model induces an AMM, it follows that long before the expiration time, within epochs, our construction satisfies all of the properties and corollaries in \cite{EngelH2021}. However, approaching the expiration $t_e$ of the expiring token, the assumptions in \cite{EngelH2021} begin to break down as market conditions shift. Thus an additional mechanism -- in particular, the market clearing mechanism in section 8 -- is needed to clear the market in the case of extreme conditions, such as when all liquidity providers are inactive or no more tokens can be sold using the AMM component of the DEX.


Now we study the behavior of price and slippage with respect to the variable $c$ and changes in liquidity states. In particular, we have the following definitions.
\begin{definition}[Price and Slippage]\label{price_and_slippage}
    Given a state curve $f_{c, x_0, y_0}$ and a particular state $(x,y) \in \state(f_{c, x_0,y_0})$, the \emph{instantaneous price}, or the \emph{spot price}, at $(x,y)$ is defined by
    \[
    p_{c, x_0, y_0}(x) = \frac{1}{-f_{c, x_0,y_0}'(x)}.
    \]
    Additionally, we define the \emph{instantaneous slippage} by
    \[
    s_{c, x_0, y_0}(x) = p_{c, x_0, y_0}'(x) = \frac{f_{c, x_0,y_0}''(x)}{(f_{c, x_0,y_0}'(x))^2}.
    \]
\end{definition}
\begin{proposition}
    Given a state $(x_0, y_0)$, the instantaneous price $p_{c, x_0, y_0}(x_0)$ and slippage $s_{c, x_0, y_0}(x_0)$ are strictly decreasing in $c$, for $c > 0$. That is, if $0 < c_1 < c_2$, then 
    \[
    p_{c_1, x_0, y_0}(x_0) > p_{c_2, x_0, y_0}(x_0) \quad \text{and} \quad  s_{c_1, x_0, y_0}(x_0) > s_{c_2, x_0, y_0}(x_0).
    \]
\end{proposition}
\begin{proof}
    Observe that for all $c > 0$,
    \[
    f_{c, x_0, y_0}'(x) = -c y_0 \left(\frac{x}{x_0}\right)^{-c-1}\cdot \frac{1}{x_0}, \qquad f_{c, x_0, y_0}''(x) = c(c+1) y_0 \left(\frac{x}{x_0}\right)^{-c-2}\cdot \frac{1}{x_0^2},
    \]
    so that
    \[
    f_{c, x_0, y_0}'(x_0) = -c\frac{y_0}{x_0}, \qquad f_{c, x_0, y_0}''(x_0) = c(c+1)\frac{y_0}{x_0^2}
    \]
    Thus
    \[
    p_{c_1, x_0, y_0}(x_0) = \frac{1}{-f_{c_1, x_0, y_0}'(x_0)} = \frac{x_0}{y_0c_1} > \frac{x_0}{y_0c_2} = \frac{1}{-f_{c_2, x_0, y_0}'(x_0)} = p_{c_2, x_0, y_0}(x_0).
    \]
    Furthermore, for all $c > 0$,
    \[
    s_{c, x_0, y_0}(x_0) = \frac{f_{c, x_0,y_0}''(x_0)}{(f_{c, x_0,y_0}'(x_0))^2} = \frac{c(c+1)\frac{y_0}{x_0^2}}{\left(-\frac{cy_0}{x_0}\right)^2} = \frac{c+1}{c}\cdot \frac{1}{y_0} = \frac{1}{y_0}\cdot \left(1 + \frac{1}{c}\right).
    \]
    Therefore
    \[
    s_{c_1, x_0, y_0}(x_0) = \frac{1}{y_0}\cdot \left(1 + \frac{1}{c_1}\right) > \frac{1}{y_0}\cdot \left(1 + \frac{1}{c_2}\right) = s_{c_2, x_0, y_0}(x_0).
    \]
\end{proof}

Finally, we prove that adding and removing liquidity does not change the instantaneous price. As discussed in the next section, we require that the ratio $x/y$ of active liquidity in the AMM stays constant during these operations. It is also insightful to prove that adding liquidity decreases instantaneous slippage, and removing liquidity increases slippage. Since adding and removing liquidity holds $x/y$ constant, the following proposition shows these results.
\begin{proposition}\label{price_invariance}
    Let $c > 0$, and consider a state $(x_1, y_1) \in \state(f_{c, x_0, y_0})$ and some $(x_2, y_2) \in \R_{>0}^2$ such that $\frac{x_1}{y_1} = \frac{x_2}{y_2}$, where we fix $(x_0, y_0) \in \R_{>0}^2$. Then $p_{c, x_0, y_0}(x_1) = p_{c, x_2, y_2}(x_2)$. Furthermore, $s_{c, x_0, y_0}(x_1) > s_{c, x_2, y_2}(x_2)$ if and only if $y_1 < y_2$.
\end{proposition}
\begin{proof}
    Observe that 
    \[
    f_{c, x_0, y_0}'(x) = -c y_0 \left( \frac{x}{x_0} \right)^{-c-1} \frac{1}{x_0} = -c \cdot \frac{f_{c, x_0, y_0}(x)}{x}
    \]
    for all $x > 0$. Because $(x_1, y_1) \in \state(f_{c, x_0,y_0})$ and $(x_2,y_2) \in \state(f_{c, x_2, y_2})$, we have
    \[
    p_{c, x_0, y_0}(x_1) = \frac{1}{c \cdot \frac{f_{c, x_0, y_0}(x_1)}{x_1}} = \frac{x_1}{cy_1} = \frac{x_2}{cy_2} = \frac{1}{c \cdot \frac{f_{c, x_2, y_2}(x_2)}{x_2}} = p_{c, x_2, y_2}(x_2).
    \]
    We also have 
    \[
    f_{c, x_0, y_0}''(x) = c(c+1) y_0 \left(\frac{x}{x_0}\right)^{-c-2}\cdot\frac{1}{x_0^2} = c(c+1) \cdot \frac{f_{c, x_0, y_0}(x)}{x^2},
    \]
    so that $s_{c, x_0, y_0}(x) = \frac{c+1}{c} \cdot \frac{1}{f_{c, x_0, y_0}(x)}$ for all $x > 0$. It follows that
    \[
    s_{c, x_0, y_0}(x_1) = \frac{c+1}{c} \cdot \frac{1}{y_1}, \qquad s_{c, x_2, y_2}(x_2) = \frac{c+1}{c} \cdot \frac{1}{y_2}
    \]
    since $(x_1, y_1) \in \state(f_{c, x_0,y_0})$ and $(x_2,y_2) \in \state(f_{c, x_2, y_2})$. Hence $s_{c, x_0, y_0}(x_1) > s_{c, x_2, y_2}(x_2)$ if and only if $y_1 < y_2$.
\end{proof}

\section{Freezing Liquidity and Related Methods}
\label{Sec-FreezingAndRelated}
With expiring assets (and exogenous events that affect price), it is important to avoid that LPs have their cash ($X$) drained upon a sudden decrease in demand for tokens ($Y$). 
In this section, we discuss \emph{liquidity freezing},
a tool allowing an LP to alter how much of its liquidity 
the AMM can use for trading under current market conditions.

LPs add liquidity in a way almost identical to AMMs such as Uniswap~\cite{uniswapv2}.
If $(x,y)$ is the current liquidity state of the AMM,
then any party $\ell$ may deposit $x_\ell> 0$ of $X$ and $y_\ell>0$ of $Y$
in a proportion that reflects the current price: $\frac{x_\ell}{y_\ell} = \frac{x}{y}$.\footnote{In an initial state where $x=0$ or $y=0$,
there are no restrictions on how liquidity is added.} Then the state $(x,y)$ is updated appropriately, we do $x_0 \gets x$ and $y_0 \gets y$, and finally, $L \gets L \cup \{\ell\}$.
The depositing LP receives shares in the usual way,
but in addition it is allowed to specify its contribution $c_\ell \in [a,b]$ in the $\mathtt{aggregate}()$ algorithm (Equation~\ref{eqn:c}),
changing the state curve at the next epoch boundary.
However, an LP can freeze its liquidity when it considers market conditions to be adverse.
The LP establishes a bound on the spot price of tokens ($Y$)
in terms of cash ($X$) below which its liquidity will not participate.
Each liquidity provider $\ell \in L \cup L'$ sets a lower bound $d_\ell \ge 0$,
updated at the beginning of each epoch (after $\mathtt{removeLiquidity}()$ and $\mathtt{liquidate}()$ defined subsequently). A liquidity provider $\ell$ can \emph{alter} its $d_\ell$ at any time, but no change will occur until the next epoch boundary. 

For any initial state $(x_0, y_0)$ and current state $(x,y)$ of the AMM with aggregate constant $c$, let $p$ denote the (instantaneous) \emph{spot price} 
\[
p = \frac{1}{-f_{c, x_0, y_0}'(x)},
\]
as defined in Definition \ref{price_and_slippage}.
This value is stored in the AMM and is updated appropriately.
As soon as $p < d_\ell$ for some $\ell$,
the AMM moves $\ell$'s liquidity into a distinct,
untraded liquidity pool.
If, at a later time, $p \ge d_\ell$,
the AMM unfreezes $\ell$'s liquidity by moving it back into the trading state.
To ensure no consumer or producer pushes $p$ to be much smaller than some $d_\ell$,
trades are restricted to \emph{one unit increments}. The assumption is primarily made for simplification of the proof of Proposition \ref{prop:remove_liquidate}. For completeness, we write the exact trading algorithm for multiple tokens $Y$ and the corresponding system state changes in pseudocode below. A more efficient implementation is a discretized version of Uniswap V3's concentrated liquidity \cite{uniswapv3}. We delay discussion of Lines~\ref{line:freeze1}-\ref{line:freeze2} and ~\ref{line:freeze3}-\ref{line:freeze4} to Section \ref{Sec-MarketClearing}.

\begin{algorithm}[!ht]\label{Alg-freeze_unfreeze}
\DontPrintSemicolon
\caption{$\mathtt{trade(tokenIn, amountIn, address)}$}
\If{$\mathtt{tokenIn} = X$}{
    $\mathtt{amountUsed} \gets 0$

    $\mathtt{done} \gets \mathtt{false}$

    \While{$L \ne \emptyset$ and not $\mathtt{done}$}{
        Let $\mathtt{price}$ be the current price in $X$ to buy one $Y$ token

        $\mathtt{amountUsed} \gets \mathtt{amountUsed} + \mathtt{price}$

        \If{$\mathtt{amountUsed} > \mathtt{amountIn}$}{
            $\mathtt{done} \gets \mathtt{true}$
        }

        \Else{
            $\mathtt{transferIn}(\mathtt{price}, X, \mathtt{source = address})$
            
            $\mathtt{transferOut}(1, Y, \mathtt{destination = address})$

            $\mathtt{freeze\_unfreeze}()$

            \For{$\ell \in L$}{\label{line:freeze1}
                $r_\ell \gets r_\ell + s_\ell \cdot \mathtt{price}$\label{line:freeze2}
            }
        }
    }
}
\If{$\mathtt{tokenIn} = Y$}{
    $\mathtt{tokensUsed} \gets 0$

    $\mathtt{done} \gets \mathtt{false}$

    \While{$L \ne \emptyset$ and not $\mathtt{done}$}{
        Let $\mathtt{price}$ be the current price in $X$ to sell one $Y$ token

        $\mathtt{tokensUsed} \gets \mathtt{tokensUsed} + 1$

        \If{$\mathtt{tokensUsed} > \mathtt{amountIn}$}{
            $\mathtt{done} \gets \mathtt{true}$
        }

        \Else{
            $\mathtt{transferIn}(1, Y, \mathtt{source = address})$
            
            $\mathtt{transferOut}(\mathtt{price}, X, \mathtt{destination = address})$

            $\mathtt{freeze\_unfreeze}()$

            \For{$\ell \in L$}{\label{line:freeze3}
                $r_\ell \gets r_\ell + s_\ell \cdot \mathtt{price}$\label{line:freeze4}
            }
        }
    }
}
\end{algorithm}


We showed in Proposition \ref{price_invariance} that the price $p$ is invariant
under adding and removing liquidity,
so we update $p$ only on epoch boundaries and after every trade.
The AMM maintains the set $\{d_\ell\}_{\ell \in L \cup L'}$ and a map $\mathtt{fr}: L' \to \R^2$ to keep track of the liquidity that inactive providers have the right to.
Immediately after every epoch boundary and every trade,
we execute the method $\mathtt{freeze\_unfreeze}()$ 
defined in Algorithm \ref{Alg-freeze_unfreeze}.

\begin{algorithm}[!ht]\label{Alg-freeze_unfreeze}
\DontPrintSemicolon
\caption{$\mathtt{freeze\_unfreeze()}$}
$\mathtt{unfreeze}()$

$\mathtt{freeze}()$
\end{algorithm}

The $\mathtt{freeze}()$ function finds all active liquidity providers $\ell$ whose price tolerance $d_\ell$ is less than the current price,
and moves their liquidity into a non-trading pool.
If $L \ne \emptyset$,
$\mathtt{unfreeze}()$ finds the inactive liquidity providers $\ell$ whose price
tolerance $d_\ell$ is at least $p$, and moves as much of their inactive liquidity back into the trading pool as possible,
subject to preserving the ratio of $x$ to $y$ in the liquidity pool.
If, however, $L = \emptyset$, the instantaneous price $p$ is not well-defined,
so we retrieve the values of $x$ and $y$ from the last point in time where some liquidity provider was active, and unfreeze all the liquidity providers to have a liquidity ratio of $x/y$. In both cases $L = \emptyset$ or $L \ne \emptyset$, the $\mathtt{freeze\_unfreeze}()$ method freezes or unfreezes liquidity providers as needed.

\SetKwInput{KwInput}{Input}
\SetKwInput{KwOutput}{Output}

\begin{algorithm}[!ht]
\DontPrintSemicolon
\caption{$\mathtt{freeze()}$}

$L_{\mathtt{freeze}} \gets \{\ell \in L: p < d_\ell\}$\\
$L \gets L \setminus L_{\mathtt{freeze}}$\\
$L' \gets L' \cup L_{\mathtt{freeze}}$\\
\For{$\ell \in L_{\mathtt{freeze}}$}{
    $\mathtt{fr}(\ell) \gets (s_\ell x, s_\ell y)$\\
}
$s_{\mathtt{total}} \gets \sum_{k \in L} s_k$\\
$x \gets s_{\mathtt{total}} x$\\
$y \gets s_{\mathtt{total}} y$\\
\For{$\ell \in L$}{
    $s_\ell \gets \frac{s_\ell}{s_{\mathtt{total}}}$
}
\end{algorithm}

\begin{algorithm}[!ht]
\DontPrintSemicolon
\caption{$\mathtt{unfreeze()}$}

\If{$L = \emptyset$}{
    Let $r$ be the ratio $x/y$ the last instant where $L \ne \emptyset$.\\
    \While{$\exists \ell \in L'$}{
        $\mathtt{unfreezeProvider}(\ell, r)$\\
    }
    $L \gets L'$\\
    $L' \gets \emptyset$\\
}
\Else{
    $L_{\mathtt{unfreeze}} \gets \{\ell \in L': p \ge d_\ell\}$\\
    \While{$\exists \ell \in L_{\mathtt{unfreeze}} \cap L'$}{
        $\mathtt{unfreezeProvider}(\ell, x/y)$
    }
}
\end{algorithm}

\begin{algorithm}[!ht]
\DontPrintSemicolon
\caption{$\mathtt{unfreezeProvider}(\ell, r)$}

\textbf{require} $\ell \in L'$

$(x_\ell, y_\ell) \gets \mathtt{fr}(\ell)$\\
\If{$x_\ell/y_\ell \ge r$}{
    $x_\ell' \gets ry_\ell$\\
    $y_\ell' \gets y_\ell$\\
    $y \gets y+y_\ell$\\
    $x \gets x + x_\ell'$\\
    $\mathtt{transferOut}(x_\ell-x_\ell', X, \mathtt{destination} = \ell)$
}
\Else{
    $x_\ell' \gets x_\ell$\\
    $y_\ell' \gets x_\ell/r$\\
    $y \gets y+y_\ell'$\\
    $x \gets x + x_\ell$\\
    $\mathtt{transferOut}(y_\ell-y_\ell', Y, \mathtt{destination} = \ell)$
}
$L' \gets L' \setminus \{\ell\}$\\
$L \gets L \cup \{\ell\}$\\
$s_\ell \gets \frac{x_\ell'}{x}$\\
\For{$l \in L \setminus \{\ell\}$}{
    $s_l \gets s_l \cdot (1-s_\ell)$
}

\end{algorithm}


Finally, we define $\mathtt{removeLiquidity}()$ and $\mathtt{liquidate}()$. These methods are similar, but with one key difference.
When $\mathtt{removeLiquidity}()$ is called by some $\ell \in L \cup L'$, the AMM adds $\ell$ to a set $L_{\mathtt{remove}}$ initialized to $\emptyset$;
at the next epoch boundary,
the AMM simply sends the assets that each $\ell \in L_{\mathtt{remove}}$ has the right to (computed in the obvious way, depending on if $\ell \in L$ or $\ell \in L'$) back to $\ell$,
while doing the appropriate bookkeeping.
As this is the only difference from usual practice
(\emph{e.g.,}~\cite{uniswapv2}),
so pseudocode is omitted.

Like $\mathtt{removeLiquidity}()$,
when $\mathtt{liquidate}()$ is called by an LP $\ell \in L \cup L'$,
then $\ell$ is added to a set $L_{\mathtt{liquidate}}$.
At each epoch boundary, the following is executed in an atomic step:
for each liquidity provider in $L_{\mathtt{liquidate}}$,
get the amounts of $X$ and $Y$ they have the right to,
remove this liquidity,
sell the total units of $Y$ that have been traded for $X$,
and distribute these funds to each member of $L_{\mathtt{liquidate}}$
in proportion to their shares of asset $X$;
then set $L_{\mathtt{liquidate}} \gets \emptyset$.
This protocol is listed as Algorithm~\ref{Alg-Liquidate}.
Both these methods run atomically at each epoch boundary,
where $\mathtt{removeLiquidity}()$ is executed first.

\begin{algorithm}[!ht]
\label{Alg-Liquidate}
\DontPrintSemicolon
\caption{$\mathtt{liquidate()}$}
\While{$\exists \ell \in L_{\mathtt{liquidate}} \cap L$}{
    $x_\ell \gets s_\ell x$\\
    $y_\ell \gets s_\ell y$\\
    $x \gets x - x_\ell$\\
    $y \gets y - y_\ell$\\
    $L \gets L \setminus \{\ell\}$
}
\While{$\exists \ell \in L_{\mathtt{liquidate}} \cap L'$}{
    $(x_\ell, y_\ell) \gets \mathtt{fr}(\ell)$\\
    $L' \gets L' \setminus \{\ell\}$
    \texttt{delete} $\mathtt{fr}(\ell)$
}
$s_\mathtt{total} \gets \sum_{k \in L} s_k$\\
\For{$\ell \in L$}{
    $s_\ell \gets \frac{s_\ell}{s_\mathtt{total}}$
}
$y_{\mathtt{liquidate}} \gets \sum_{\ell \in L_{\mathtt{liquidate}}} y_\ell$

$x_{\mathtt{liquidate}} \gets 0$

\For{$\ell \in L_{\mathtt{liquidate}}$}{
    $\overline{s_\ell} \gets \frac{y_\ell}{y_{\mathtt{liquidate}}}$
}

/* Sell the $y_{\mathtt{liquidate}}$ units of $Y$ back to the AMM one at a time */

\While{$y_{\mathtt{liquidate}} > 0$ and $L \ne \emptyset$}{
    Trade one unit of $Y$ for $x_{\mathtt{returned}}$ units of $X$ \hfill /* This updates state\\\hfill variables appropriately */

    $x_{\mathtt{liquidate}} \gets x_{\mathtt{liquidate}} + x_{\mathtt{returned}}$\\
    $y_{\mathtt{liquidate}} \gets y_{\mathtt{liquidate}} - 1$
}

\For{$\ell \in L_{\mathtt{liquidate}}$}{
    $x_\ell \gets x_\ell + \overline{s_\ell} \cdot x_{\mathtt{liquidate}}$\\
    $y_\ell \gets \overline{s_\ell} \cdot y_{\mathtt{liquidate}}$\\
    $\mathtt{transferOut}(x_\ell, X, \mathtt{destination} = \ell)$\\
    $\mathtt{transferOut}(y_\ell, Y, \mathtt{destination} = \ell)$
}
$L_{\mathtt{liquidate}} \gets \emptyset$
\end{algorithm}



\subsection{Freezing, Removal, and Liquidation Properties}
The advantage of including two different ways of removing liquidity and only at epoch boundaries is primarily to spread out the loss of the liquidity providers who choose to liquidate and provide those who call $\mathtt{removeLiquidity}()$ a guaranteed risk-free liquidity removal; the received liquidity from the $\mathtt{removeLiquidity}()$ method is not affected by liquidity providers calling $\mathtt{liquidate}()$ since the AMM's execution of $\mathtt{removeLiquidity}()$ precedes that of $\mathtt{liquidate}()$. 

We will prove a more formal version of the following claim: no rational liquidity provider will call $\mathtt{removeLiquidity}()$ and then sell their $Y$ tokens back to the AMM in exchange for $X$. In particular, this means that liquidity providers who wish to keep their $Y$ tokens upon removal call the $\mathtt{removeLiquidity}()$ method, and liquidity providers who only want to remove liquidity with the maximum number of $X$ tokens possible ($Y$ is worthless to them) call the $\mathtt{liquidate}()$ method. First, we prove the following lemmas.

\begin{lemma}\label{price_bound}
    Let $c > 0$ and $(x_0, y_0) \in \R_{>0}^2$. If $(x', y') \in \state(f_{c, x_0, y_0})$ and $(x'', y'') \in \R_{>0}^2$ satisfy $\frac{x'}{y'} = \frac{x''}{y''}$ and $x'' < x'$ and $y'' < y'$, then 
    \[
    |f_{c, x'', y''}'(x)| \ge |f_{c, x_0, y_0}'(x+x'-x'')|,
    \]
    for all $x \in (0, x'']$.
\end{lemma}
\begin{proof}
    Suppose $(x', y') \in \state(f_{c, x_0, y_0})$ and $(x'', y'') \in \R_{>0}^2$ satisfy $\frac{x'}{y'} = \frac{x''}{y''}$ and $x'' < x'$ and $y'' < y'$. Notice that
    \[
    f_{c, x_0, y_0}'(x) = -c y_0 \left( \frac{x}{x_0} \right)^{-c-1} \cdot \frac{1}{x_0} = -c \cdot \frac{f_{c, x_0, y_0}(x)}{x},
    \]
    and similarly, 
    \[
    f_{c, x'', y''}'(x) = -c \cdot \frac{f_{c, x'', y''}(x)}{x}.
    \]
    Since $(x', y') \in \state(f_{c, x_0, y_0})$, we know 
    \[
    f_{c, x_0, y_0}(x) = y_0 x_0^c x^{-c} = y_0 \left( \frac{x'}{x_0} \right)^{-c} (x')^c x^{-c} = f_{c, x_0, y_0}(x') \cdot (x')^c x^{-c} = y' (x')^c x^{-c}.
    \]
    Fix some $x \in (0, x'']$. Since $x' - x'' \ge 0$, $0 < x'' \le x'$, and $\frac{x'}{x''} = \frac{y'}{y''}$, we have
    \begin{align*}
    \left( \frac{x+x'-x''}{x} \right)^{c+1} &\ge \left( 1 + \frac{x'-x''}{x} \right)^{c+1} \ge \left( 1 + \frac{x'-x''}{x''} \right)^{c+1}\\
    &= \left(\frac{x'}{x''}\right)^{c+1} = \frac{y'}{y''} \cdot \left(\frac{x'}{x''}\right)^c.
    \end{align*}
    It follows that
    \[
    y'' \cdot x^{-c-1} (x'')^c \ge y' (x+x'-x'')^{-c-1} (x')^c.
    \]
    Multiplying by $c$ yields that
    \begin{align*}
    |f_{c, x'', y''}'(x)| &= c \cdot \frac{f_{c, x'', y''}(x)}{x} = c \cdot y'' x^{-c-1} (x'')^c\\
    &\ge c \cdot \frac{y'\left(\frac{x+x'-x''}{x'}\right)^{-c}}{x+x'-x''} = c \cdot \frac{f_{c, x_0, y_0}(x+x'-x'')}{x+x'-x''}\\
    &= |f_{c, x_0, y_0}'(x+x'-x'')|.
    \end{align*}
\end{proof}

\begin{lemma}\label{sell-convexity}
    Suppose some party sells one unit of $Y$ to the AMM for $x_1$ of $X$, and then sells another unit of $Y$ to the AMM for $x_2$ of $X$. Then $x_1 \ge x_2$.
\end{lemma}
\begin{proof}
    Let $f_{c, x_0,y_0}$ be the state curve before the first trade. If $L$ and $L'$ are the same after the first trade (so $\mathtt{freeze\_unfreeze}()$ did not update any global variables), then since $f_{c, x_0, y_0}$ is a convex function, we certainly have $x_1 \ge x_2$.

    Now suppose the execution of $\mathtt{freeze\_unfreeze}()$ changed the sets $L$ and $L'$ after the first trade. It suffices to prove that $x_2$ is at most the amount of $X$ one would get if the state curve never updated and was always equal to $f_{c, x_0, y_0}$, given the observation above. To this end, we consider two possible cases for the second trade. Let $(x', y') \in \state(f_{c, x_0, y_0})$ be the state after the first trade but before calling $\mathtt{freeze\_unfreeze}()$; let $(x'', y'')$ be the state after completing the first trade, so the state curve updates to $f_{c, x'', y''}$. Observe that the updated set $L$ after the first trade is a subset of the initial $L$ because the instantaneous price $p$ can only decrease after the first selling trade. By inspection of $\mathtt{freeze}()$ and $\mathtt{unfreeze}()$, we see that $\frac{x'}{y'} = \frac{x''}{y''}$ but $0 < x'' < x'$ and $0 < y'' < y'$.
    
    Let $x_2'$ be the amount received for the second trade if the state curve is $f_{c, x_0, y_0}$; let $x_2''$ be the amount received for the second trade if the state curve is $f_{c, x'', y''}$. The convexity of $f_{c, x_0, y_0}$ implies that $x_1 \ge x_2'$. If we can show $x_2' \ge x_2''$, then we know $x_1 \ge x_2$ because $x_2 \in \{x_2', x_2''\}$.

    By Lemma \ref{price_bound}, $|f_{c, x'', y''}'(x)| \ge |f_{c, x_0, y_0}'(x+x'-x'')|$ for all $x \in (0, x'']$. This implies that
    \[
    \left| \frac{d}{dy} f^{-1}_{c, x'', y''}(y) \right| \le \left| \frac{d}{dy} f^{-1}_{c, x_0, y_0}(y+y'-y'') \right|
    \]
    for all $y \in [y'', \infty)$.

    Finally, we prove $x_2' \ge x_2''$, showing the lemma. We have
    \begin{align*}
        x_2' &= f^{-1}_{c, x_0, y_0}(y') - f^{-1}_{c, x_0, y_0}(y'+1)\\
        &= \int_{y''+1}^{y''} \frac{d}{dy} f^{-1}_{c, x_0, y_0}(y+y'-y'') dy\\
        &= \int_{y''}^{y''+1} \left| \frac{d}{dy} f^{-1}_{c, x_0, y_0}(y+y'-y'') \right| dy\\
        &\ge \int_{y''}^{y''+1} \left| \frac{d}{dy} f^{-1}_{c, x'', y''}(y) \right| dy\\
        &= \int_{y''+1}^{y''} \frac{d}{dy} f^{-1}_{c, x'', y''}(y) dy\\
        &= f^{-1}_{c, x'', y''}(y'') - f^{-1}_{c, x'', y''}(y''+1) = x_2''.
    \end{align*}
\end{proof}

\begin{proposition}\label{prop:remove_liquidate}
    Let $\ell \in L \cup L'$. Suppose that in some execution $\alpha$, the liquidity provider $\ell$ calls $\mathtt{removeLiquidity}()$ to obtain $x^\alpha_{\ell,1}$ of $X$ and $y^\alpha_{\ell,1}$ of $Y$, and then trades $y^\alpha_{\ell,1}$ of $Y$ to the AMM in exchange for $x^\alpha_{\ell, 2}$ of $X$, so that $\ell$ receives $x^\alpha_\ell = x^\alpha_{\ell,1} + x^\alpha_{\ell,2}$ of $X$. Suppose that in some otherwise identical execution $\beta$, $\ell$ instead calls $\mathtt{liquidate}()$ and receives $x^\beta_{\ell}$ of $X$. Then $x_\ell^\alpha \le x_\ell^\beta$.
\end{proposition}
\begin{proof}
    Consider the sets $L_{\mathtt{liquidate}}^\alpha$ and $L_{\mathtt{liquidate}}^\beta$ in executions $\alpha$ and $\beta$, respectively. Consider the $c$, which is the same in both executions. Notice that, by construction, the effective AMM state $(x_0,y_0)$ after executing the loop in line 14 of $\mathtt{liquidate}()$ is the same in both executions $\alpha$ and $\beta$. Suppose that in execution $\alpha$, there is some other liquidity provider (not $\ell$) that removed liquidity and sold it to the system before $\ell$. By Lemma \ref{sell-convexity}, the received amount $x^\alpha_{\ell, 2}$ of $X$ is at most the amount $\ell$ would have received if $\ell$ was the first provider to sell their removed liquidity back to the AMM. Thus it suffices to prove the claim when $\ell$ was the first liquidity provider to sell their liquidity back to the AMM after $\mathtt{liquidate}()$ was called in execution $\alpha$.
    
    To this end, consider the sequence $(x^\alpha_k)_{k=1}^n$ of received amounts of $X$ for consecutive single-unit trades made by $\mathtt{liquidate}()$ and $\ell$'s atomic selling trades, in execution $\alpha$. Similarly, let $(x^\beta_k)_{k=1}^n$ be the sequence of received amounts of $X$ for consecutive single-unit trades made by $\mathtt{liquidate}()$, in execution $\beta$. It is clear that $n \le y_\mathtt{liquidate}$, as defined in $\mathtt{liquidate}()$, so that we may extend both sequences by setting $x^\alpha_k = x^\beta_k = 0$ for $n < k \le y_\mathtt{liquidate}$ By the assumption made above, these two sequences are equal: $x^\alpha_k = x^\beta_k$ for all $k \in [y_\mathtt{liquidate}]$. Then we may let $(x_k)_{k=1}^{y_\mathtt{liquidate}}$ be this sequence. By Lemma \ref{sell-convexity}, $(x_k)_{k=1}^{y_\mathtt{liquidate}}$ is a nonincreasing sequence. Thus
    \[
    x^\alpha_{\ell, 2} = \sum_{k = y_\mathtt{liquidate}-y^\alpha_{\ell,1}+1}^{y_\mathtt{liquidate}} x^\alpha_k \le \frac{y^\alpha_{\ell,1}}{y_{\mathtt{liquidate}}} \cdot \sum_{k=1}^{y_\mathtt{liquidate}} x_k = \overline{s_\ell} \cdot x_\mathtt{liquidate},
    \]
    where $\overline{s_\ell}$ and $x_\mathtt{liquidate}$ are defined in $\mathtt{liquidate}()$. Note that the $x_\ell$ in $\mathtt{liquidate}()$ is equal to $x_{\ell,1}^\alpha$, so that
    \[
    x_\ell^\alpha = x_{\ell,1}^\alpha + x_{\ell, 2}^\alpha \le x_{\ell, 1}^\alpha + \overline{s_\ell} \cdot x_\mathtt{liquidate} = x_\ell^\beta.
    \]
\end{proof}

We also show that the $\mathtt{freeze\_unfreeze}()$ method yields a \emph{stable} liquidity state; that is, a liquidity provider $\ell \in L \cup L'$ is active if and only if $p \ge d_\ell$. Proving this is important for proving LP Stability in Theorem \ref{market_clearing}.
\begin{proposition}\label{LP_stability}
    Consider the AMM state immediately after calling $\mathtt{freeze\_unfreeze}()$. Let $\ell \in L \cup L'$. Then $\ell \in L$ if and only if $p \ge d_\ell$.
\end{proposition}
\begin{proof}
    Inspecting both $\mathtt{freeze}()$ and $\mathtt{unfreeze}()$, the AMM state changes of the variables $(x,y)$ are indistinguishable from a sequence of adding and removing liquidity. By Lemma \ref{price_invariance}, the current instantaneous price $p$ is invariant throughout $\mathtt{freeze\_unfreeze}()$. If $L = \emptyset$ before executing $\mathtt{freeze\_unfreeze}()$, then this implies that after executing $\mathtt{unfreeze}()$ and $\mathtt{freeze}()$, we can see that $\ell \in L$ if and only if $\ell \notin L_{\mathtt{freeze}}$ (as defined in $\mathtt{freeze}()$), which holds if and only if $p \ge d_\ell$. On the other hand, if $L \ne \emptyset$ before executing $\mathtt{freeze\_unfreeze}()$, then by a similar reasoning, $\ell \in L$ after executing the algorithm if and only $p \ge d_\ell$.
\end{proof}

\section{Market Clearing Auction}
\label{Sec-MarketClearing}
In this section, we describe a mechanism for bargain hunters, 
consumers with low price and low urgency willing to wait until the last minute to acquire tokens. We also describe a complementary mechanism for producers to ensure market clearance.

The first protocol we use is a form of a second-price auction.
Importantly, we fairly distribute these profits to LPs according to their cumulative participation in the protocol before retrieval time $t_r$. Informally, if an LP $\ell$ takes on more risk by participating in a large number of trades with a large share of liquidity, then, in addition to receiving more fees, the market clearing auction will prioritize clearing $\ell$'s remaining tokens ($Y$) for cash ($X$) at a higher price than other, more risk-averse, LPs.

A \emph{consumer order}, or \emph{bid} $b$ has two attributes:
\begin{itemize}
    \item $\mathtt{address}(b)$: if a party $u$ submits the order $b$, we set $\mathtt{address}(b) = u$.
    \item $\mathtt{price}(b)$: the party $\mathtt{address}(b)$ sets a price $\mathtt{price}(b) > 0$ that they are willing to pay for one unit of the token.
\end{itemize}
When a bid is submitted, it is not locked; any user can remove any of their bids if desired, at any time.
For ease of exposition,
each bid is for a single unit of $Y$. A user who wishes to buy multiple units of $Y$ simply submits multiple bids.

The AMM maintains a dynamic list $\mathcal{B}$ that keeps track of all existing bids at any given time. The list $\mathcal{B}$ is called the \emph{consumer order book}. Initially $\mathcal{B}$ is empty, and every time a bid $b$ is submitted, $b$ is added to $\mathcal{B}$, so that $\mathcal{B}$ is monotone decreasing by the price of each bid (highest price to lowest price). This implies that $\mathcal{B}[0]$ is a bid with maximum price. If the AMM price $p_Y$ for a single unit of $Y$ falls to below $\mathtt{price}(b)$ for the bid $b = \mathcal{B}[0]$, then $b$ is removed from $\mathcal{B}$ and one unit of $Y$ is immediately sold to $\mathtt{address}(b)$ via a trade at the price $p_Y \le \mathtt{price}(b)$, so that the profits and fees of the trade are distributed to the active liquidity providers ($L$) in proportion to their shares as usual.

We next describe the market clearing algorithm $\mathtt{clearing}()$. From the perspective of a buyer, $\mathtt{clearing}()$ is purely a second-price auction. However, the distribution of profits to the LPs is more intricate. Each LP $\ell \in L \cup L'$ is assigned some variable $r_\ell$ that informally keeps track of the cumulative risk $\ell$ has faced.\footnote{$r_\ell$ is initialized to $0$ when $\ell$ joins the set of LPs.} When a trade (buying or selling) that exchanges $x'$ of $X$ and some amount of $Y$, we do the following:
\begin{itemize}
    \item For each $\ell \in L$, compute $x_\ell \gets s_\ell x'$ and increment $r_\ell \gets r_\ell + x_\ell$.\footnote{Fees are also distributed in proportion to this $x_\ell$.}
\end{itemize}

Now consider the state of the AMM at the retrieval time $t_r$. We execute the following market clearing algorithm $\mathtt{clearing}()$ at this time.

\begin{algorithm}[!ht]
\DontPrintSemicolon
\caption{$\mathtt{clearing()}$}
\For{$\ell \in L$}{
    $\mathtt{Global}: x_\ell \gets s_\ell x$\\
    $\mathtt{Global}: y_\ell \gets s_\ell y$
}
\For{$\ell \in L'$}{
    $\mathtt{Global}: (x_\ell, y_\ell) \gets \mathtt{fr}(\ell)$
}
\For{$\ell \in L \cup L'$}{
    $\mathtt{Global}: \overline{s_\ell} \gets \frac{r_\ell}{\sum_{k \in L \cup L'} r_k}$
}

$\mathtt{Global}: L_{\mathtt{available}} \gets L \cup L'$

\While{$|\mathcal{B}| \ge 1$ and $\sum_{\ell \in L \cup L'} y_\ell \ge 1$}{
    $b \gets \mathcal{B}[0]$
    
    $\mathtt{delete}$ $\mathcal{B}[0]$

    $p \gets \mathtt{price}(\mathcal{B}[0])$ \hfill /* Second price */

    $\mathtt{execute}(b, p)$
}
\For{$\ell \in L \cup L'$}{
    $\mathtt{transferOut}(x_\ell, X, \mathtt{destination} = \ell)$
    
    $\mathtt{transferOut}(y_\ell, Y, \mathtt{destination} = \ell)$
}
\end{algorithm}

\begin{algorithm}[!ht]
\DontPrintSemicolon
\caption{$\mathtt{execute}(b, p)$}
    $y_{\mathtt{total}} \gets 0$

    \While{$y_\mathtt{total} < 1$}{
        $L_\mathtt{toRemove} \gets \emptyset$

        \For{$\ell \in L_\mathtt{available}$}{
            \If{$y_\ell < \overline{s_\ell}(1-y_\mathtt{total})$}{
                $y_\mathtt{total} \gets y_\mathtt{total} + y_\ell$

                $y_\ell \gets 0$

                $x_\ell \gets x_\ell + p\cdot y_\ell$

                $L_\mathtt{toRemove} \gets L_{\mathtt{toRemove}} \cup \{\ell\}$

                \For{$k \in L_\mathtt{available} \setminus L_\mathtt{toRemove}$}{
                    $\overline{s_k}\gets \frac{\overline{s_k}}{1-\overline{s_\ell}}$
                }
            }
            \Else{
                $y_\mathtt{total} \gets y_\mathtt{total} + \overline{s_\ell}(1-y_\mathtt{total})$

                $y_\ell \gets y_\ell - \overline{s_\ell}(1-y_\mathtt{total})$

                $x_\ell \gets x_\ell + p\cdot \overline{s_\ell}(1-y_\mathtt{total})$
            }
        }
        $L_\mathtt{available} \gets L_\mathtt{available}\setminus L_{\mathtt{toRemove}}$
    }
    $\mathtt{transferOut}(1, Y, \mathtt{destination} = \mathtt{address}(b))$

    $\mathtt{transferIn}(p, X, \mathtt{source} = \mathtt{address}(b))$


            


            





\end{algorithm}

Essentially, we execute the bid $b$ with highest price for the second-highest price in $\mathcal{B}$, and the profits and fees from this trade are distributed so that the liquidity providers $\ell$ get priority proportional to their cumulative risk variable $r_\ell$. If all LPs $\ell \in L \cup L'$ have sufficiently many tokens, in particular, $y_\ell \ge \overline{s_\ell}$, then the profits to $\ell$ are distributed in exact proportion to $r_\ell$. Otherwise, we keep running a similar process until the system can ``complete" a unit of $Y$, keeping profits to LPs in proportion to $r_\ell$.
This provides an incentive for liquidity providers to keep the system \emph{live}: to not leave the AMM early, and to keep providing ample liquidity until the retrieval time.

In addition to maintaining the consumer order book, the AMM keeps track of a \emph{producer order book} $\mathcal{P}$, so that any party that has a unit of $Y$ can escrow it into the system, and sell it for some minimum price that they set. In particular, a \emph{producer order} $s$ has two attributes:
\begin{itemize}
    \item $\mathtt{address}(s)$: if a party $u$ submits the selling order $s$, we set $\mathtt{address}(s) = u$.
    \item $\mathtt{price}(s)$: the party $\mathtt{address}(s)$ sets a price $\mathtt{price}(s) > 0$ for which they are willing to sell one unit of the token.
\end{itemize}

The producer order book $\mathcal{P}$ is a list, initialized to $\emptyset$, that is always sorted from least to greatest (the reverse of $\mathcal{B}$) by the price of selling orders. At any time, a user can submit a producer order $s$ by escrowing one unit of $Y$ into the AMM and setting a minimum selling price $\mathtt{price}(s)$ for that producer order. If the AMM price $q_Y$ to sell one unit of $Y$ ever rises so that $s = \mathcal{P}[0]$, the producer order with the lowest price, has a price $\mathtt{price}(s) \le q_Y$, then $s$ is removed from $\mathcal{P}$ and one unit of $Y$ is immediately sold to the AMM from $\mathtt{address}(s)$ via a regular AMM trade at the price $q_Y$. 

We also impose a standard kind of interaction between $\mathcal{B}$ and $\mathcal{P}$ that functions identically to a limit order book. Any time a new consumer or producer order is submitted to the AMM, we run the following algorithm. While $|\mathcal{B}| > 0$ and $|\mathcal{P}| > 0$ and $\mathtt{price}(\mathcal{B}[0]) \ge \mathtt{price}(\mathcal{P}[0])$, then do $b \gets \mathcal{B}[0]$, $s \gets \mathcal{P}[0]$; do 
\begin{align*}
    &\mathtt{transferOut}(1, Y, \mathtt{destination} = \mathtt{address}(b));\\ 
    &\mathtt{transferOut}(\mathtt{price}(b), X, \mathtt{destination} = \mathtt{address}(s));
\end{align*}
finally, remove $b$ from $\mathcal{B}$ and remove $s$ from $\mathcal{P}$. No state variables are updated, since the funds for the orders are initially escrowed from the addresses of the orders.

This order-book style protocol is only used in extreme market conditions: (i) all but one unit of $Y$ remains in the liquidity pool, so that no consumer can make a trade with the AMM state curve; or (ii) all liquidity providers have been frozen, so that, again, no more trades can happen. In these two scenarios, the consumers and producers must submit consumer orders and producer orders, so that the interaction between the consumer order book and producer order book allows the market to remain live until the retrieval time $t_r$.






\section{Liveness and Market Clearance}
\label{Sec-LivenessClearance}
Two of the most important properties that a DEX can satisfy are \emph{liveness} and \emph{market clearance}. The liveness property in our context states that at any time before the retrieval time $t_r$, (a) there are always actions that parties can take (via the DEX) to execute a trade (trades are always possible), and (b) if the time is less than $\max T$, then all liquidity providers $\ell$ can take some action so that $\ell \in L$ after the next time in $T$ (liquidity providers are not permanently locked).

In addition to liveness, \emph{market clearance} is a desirable property of a DEX, and mirrors the safety condition usually desired in a distributed system. In this context, \emph{market clearance} means that (a) liquidity providers are always in a \emph{stable} state (in the context of Proposition \ref{LP_stability}), (b) no two parties that are willing to trade do not end up trading by retrieval time, and (c) if the DEX contains any tokens left in the system after executing the clearing protocol $\mathtt{clearing}()$, then there is no remaining consumer. Just as \emph{safety} in a distributed system requires that no correct processors ever have inconsistent views, market clearance in our AMM requires that there are no inconsistencies between the actions that parties wish to take and the transactions that actually happen in the DEX.\\

Now, we formally define and prove the liveness of our DEX model. Let $\mathtt{DynamicAMM}$ be the DEX described in this paper. We assume that there are three types of parties: buyers (consumers), sellers (producers), and liquidity providers.
\begin{definition}
    The DEX $\mathtt{DynamicAMM}$ is \emph{live} if the following hold:
    \begin{itemize}
        \item \emph{(Consumer Liveness)} Consider any time $t < t_r$, a buyer $u_b$ at time $t$, and a seller $u_s$ at time $t$. Then there is a sequence of transactions that can submitted and executed by a nonempty subset of the parties $\{u_b, u_s\}$ where $u_s$ sells a $Y$ token or $u_b$ buys a $Y$ token.
        \item \emph{(LP Liveness)} Given any liquidity provider $\ell$, a time $t < \max T$, and the AMM state at time $t$, there is a transaction that can be submitted by $\ell$ so that $\ell \in L$ when the time is at least $\min\{t' \in T: t' > t\}$ and at most $t_r$.
    \end{itemize}
\end{definition}
\begin{theorem}[Liveness]\label{liveness}
    The DEX $\mathtt{DynamicAMM}$ is live.
\end{theorem}
\begin{proof}
    Consider any time $t < t_r$, a buyer $u_b$, and a seller $u_s$. Regardless of if a trade can be made using the AMM curve of $\mathtt{DynamicAMM}$ or not, $u_b$ and $u_s$ can submit the following transactions. First, $u_b$ submits a consumer order $b$ for a price $q$; then $u_s$ submits a producer order $s$ for the same price $q$. If either of these two orders can be executed using the AMM curve, then at least one of them will be, so that we are done. Otherwise, the interaction between $\mathcal{B}$ and $\mathcal{P}$ described in Section \ref{Sec-MarketClearing}, implies that at least one of the orders $b$ and $s$ will be executed, showing that the AMM satisfies Consumer Liveness. (This still holds even if some other order is submitted between when the orders $b$ and $s$ are submitted, or if some other orders are removed.)
    
    Let $\ell$ be a liquidity provider at time $t < \max T$. Then $\ell$ can simply request that $d_\ell \gets 0$ at time $t$. Then $\mathtt{DynamicAMM}$ will update $d_\ell \gets 0$ at the time $t' = \min\{t' \in T: t' > t\}$ and then immediately run $\mathtt{freeze\_unfreeze}()$. 
    By inspection, $\ell \in L$ after running $\mathtt{unfreeze}()$. Since $p > 0$ and $d_\ell = 0$ at time $t'$, we have $p \ge d_\ell$, so that $\ell \notin L_{\mathtt{freeze}}$ in the $\mathtt{freeze}()$ algorithm; hence $\ell \in L$ after executing $\mathtt{freeze\_unfreeze}()$. At any time in $(t', t_r)$, we will still have $p > 0$, so that $\ell \in L$. Thus $\mathtt{DynamicAMM}$ also satisfies LP Liveness.
\end{proof}

Let us now formally define and prove market clearance in the context of our DEX. We assume that participants of each type are rational, as defined below:
\begin{itemize}
    \item Each liquidity provider $\ell$ has a sequence of prices\footnote{$\ell$ discovers this sequence as time progresses and is not known a priory.} $(q_t)_{t \in T} = (q(\ell)_t)_{t \in T}$ where they are willing to provide liquidity to $\mathtt{DynamicAMM}$ and engage in a trade as long as the current spot price is at least $q_{t}$, where $t$ is the greatest time in $T$ less than the current time. In this case, we assume that each $\ell \in L \cup L'$ has sets $d_\ell \gets q_t$ at each time $t \in T$. Furthermore, if the current time $t$ is in $(\max T, t_r)$, then every remaining liquidity provider is willing to sell the remaining tokens of $Y$ for any price when the time is in $(\max T, t_r)$; that is, $q_{\max T} = 0$.
    
    \item For each buyer $u_b$, there is some fixed price $q = q(u_b)$ such that if the AMM price to buy one unit of $Y$ falls to at most $q$, then $u_b$ will buy at the AMM price; if the producer order $s = \mathcal{P}[0]$ satisfies $\mathtt{price}(s) \le q$, then $u_b$ will buy from the seller $\mathtt{address}(s)$ for the price $\mathtt{price}(s)$ (by issuing a consumer order for a price $\mathtt{price}(s)$); if both of these happen at the same time, $u_b$ will buy the token $Y$ at the lower price of the two available options. At some time $t < t_r$, if $u_b$ has not bought a $Y$ token by time $t$, they will register a bid for a price $q$.
    
    \item For each seller $u_s$, there is some fixed price $q = q(u_s)$ such that if the AMM price to sell one unit of $Y$ rises to at least $q$, then $u_s$ will sell at the AMM price; if the bid $b = \mathcal{B}[0]$ with $\mathtt{price}(b) \ge q$, then $u_s$ will sell to the buyer $\mathtt{address}(b)$ for the price $\mathtt{price}(b)$; if both of these happen at the same time, $u_s$ sells the token $Y$ at the higher price of the two available options. At some time $t < t_r$, if $u_s$ has not sold a $Y$ token by time $t$, they will register a producer order for a price $q$.
\end{itemize}
\begin{definition}
    We say that $\mathtt{DynamicAMM}$ satisfies \emph{market clearance} if the following hold:
    \begin{itemize}
        \item \emph{(LP Stability)} At all times $t$ between the initialization of the AMM and retrieval time for which $L \ne \emptyset$, a liquidity provider $\ell \in L \cup L'$ is active (in $L$) if and only if $q(\ell)_{t_0} \ge p$, where $t_0$ is the greatest time in $T$ less than $t$.
        \item \emph{(Consumer Clearance)} At time $t_r$, there is no buyer $u_b$ and seller $u_s$ where $u_b$ is willing to buy at a price that is at least what $u_s$ is willing to sell for.
        \item \emph{(System Clearance)} At time $t_r$, after $\mathtt{clearing}()$ has executed, if $\mathtt{DynamicAMM}$ contains any of $Y$, then there is no remaining buyer.
    \end{itemize}
\end{definition}

\begin{theorem}\label{market_clearing}
    The DEX $\mathtt{DynamicAMM}$ satisfies market clearance.
\end{theorem}
\begin{proof}
    Consider the state of $\mathtt{DynamicAMM}$ at an arbitrary time $t$ for which $L \ne \emptyset$. Let $t_0 = \max\{t' \in T: t' < t\}$. Let $\ell \in L \cup L'$. Then by assumption, at time $t_0$, $\ell$ must have set $d_\ell \gets q(\ell)_{t_0}$. Notice that the value of $p$ at time $t$ is the same as it was at the latest time $t_1 < t$ for which $\mathtt{freeze\_unfreeze}()$ was called, since $\mathtt{freeze\_unfreeze}()$ is called precisely when $p$ changes. Then Proposition \ref{LP_stability} implies that at time $t_1$, we know $\ell \in L$ if and only if $q(\ell)_{t_0} = d_\ell \ge p$ at time $t_1$. Since $p$ is the same at time $t$ as it was at time $t_1$, we know that $\ell \in L$ if and only if $q(\ell)_{t_0} \ge p$ at time $t$. Thus $\mathtt{DynamicAMM}$ satisfies LP Stability.

    For the remainder of this proof, consider the state of $\mathtt{DynamicAMM}$ at time $t_r$.
    Suppose there is a buyer $u_b$ and a seller $u_s$ where $q(u_b) \ge q(u_s)$. Since $u_b$ and $u_s$ are unfulfilled at retrieval time, our hypothesises on buyers and sellers imply that there is a bid $b \in \mathcal{B}$ and producer order $s \in \mathcal{P}$ such that $\mathtt{address}(b) = u_b, \mathtt{address}(s) = u_s, \mathtt{price}(b) = q(u_b), \mathtt{price}(s) = q(u_s)$. If $s$ was submitted to $\mathtt{DynamicAMM}$ before $b$, then the submission of $b$ (which was before $t_r$) succeeds the check $\mathtt{price}(\mathcal{P}[0]) > \mathtt{price}(b)$ (otherwise $u_b$ would not be unfulfilled); because $\mathcal{P}$ is sorted in increasing order by price, $\mathtt{price}(s) \ge \mathtt{price}(\mathcal{P}[0]) > \mathtt{price}(b)$, so that $q(u_s) > q(u_b)$, a contradiction. The case when $b$ was submitted before $s$ gives a similar contradiction. Hence $\mathtt{DynamicAMM}$ satisfies Consumer Clearance.

    Finally, suppose that $\mathtt{DynamicAMM}$ contains at least one unit of $Y$ after $\mathtt{clearing}()$ executed, and that there is some unfulfilled buyer $u_b$. Clearly $\mathtt{clearing}()$ does eventually terminate, and $\sum_{\ell \in L \cup L'} y_\ell > 0$ when $\mathtt{clearing}()$ terminates. The termination criteria for $\mathtt{clearing}()$ imply that $\mathcal{B} = \emptyset$ after $\mathtt{clearing}()$ terminates. By our initial assumptions, $u_b$ must have submitted a consumer order for the price $q(u_b)$ before the time $t_r$, so that before $\mathtt{clearing}()$ was executed, there was some $b \in \mathcal{B}$ with $\mathtt{address}(b) = u_b$. We assumed that $u_b$ is a remaining buyer after $\mathtt{clearing}()$ was executed, which implies $b \in \mathcal{B}$ upon termination, a contradiction. Hence $\mathtt{DynamicAMM}$ satisfies System Clearance.
\end{proof}


\section{Conclusion}
This paper proposed novel DEX for trading assets that expire.
This DEX includes both AMM and auction mechanisms:
an AMM for normal execution, when expiration is far away,
and auctions when expiration is imminent.
To address market failures that may accompany expiration,
the DEX include an LP-controlled price adjustment protocol,
and a hybrid market clearing protocol that combines
the benefits of AMMs and incentive-compatible auctions.

Allowing LPs to dynamically adjust prices requires admitting a larger
family of state curves than the constant-product curves commonly
used by popular AMMs~\cite{uniswapv2, uniswapv3}.
LPs can intervene to adjust spot prices,
where an LP's influence is proportional to its investment.

As expiration approaches,
prices may become highly volatile.
To discourage LPs from withdrawing their investments,
the DEX allows nervous LPs to freeze their assets if spot
prices fall below a specified minimum,
and unfreeze if it later rises above that minimum.

The DEX includes a market clearing mechanism that functions in two ways. First, it can function as a limit order book before the expiration date to ensure market-liveness when all the liquidity providers are frozen, or when no more tokens can be bought from the system at a certain time. Second, it acts like a second-price auction to sell the remaining tokens in the system to unfulfilled customers. One key difference between our market clearing protocol and a standard second price auction is the way profits are distributed to liquidity providers: 
We distribute profits in proportion to the total participation of the liquidity providers, measured in the sum of the absolute change of the LP's cash over all trades.

We prove a number of properties which show our DEX construction is sound. First, we prove some soundness properties of our price and curve adjustment protocol -- such as monotonicity of instantaneous price as a function of $c$ and price-invariance under dynamic liquidity. Second, we prove a bound on the profit obtained by liquidity providers when given the option of calling two different, intricately connected methods of removal; this shows that it is always best to liquidate and share the losses with everyone else who wishes to liquidate, instead of attempting to act selfishly. Finally, we prove that our system is live -- which roughly means that it cannot be permanently frozen until retrieval time -- and it clears the market -- there are no two remaining parties who would have been willing to make a trade but did not meet through the system.

\bibliographystyle{plain}
\bibliography{bibliography,zotero}

\begin{thebibliography}{10}

\bibitem{uniswapv2}
Hayden Adams, Noah Zinsmeister, and Dan Robinson.
\newblock Uniswap v2 core.
\newblock \url{https://uniswap.org/whitepaper.pdf}, March 2020.
\newblock As of 8 February 2021.

\bibitem{AngerisC2020}
Guillermo Angeris and Tarun Chitra.
\newblock Improved {Price} {Oracles}: {Constant} {Function} {Market} {Makers}.
\newblock {\em SSRN Electronic Journal}, 2020.

\bibitem{AngerisEC2020}
Guillermo Angeris, Alex Evans, and Tarun Chitra.
\newblock When does the tail wag the dog? {Curvature} and market making.
\newblock {\em arXiv:2012.08040 [q-fin]}, December 2020.
\newblock arXiv: 2012.08040.

\bibitem{AngerisKCNC2019}
Guillermo Angeris, Hsien-Tang Kao, Rei Chiang, Charlie Noyes, and Tarun Chitra.
\newblock An analysis of {Uniswap} markets.
\newblock {\em arXiv:1911.03380 [cs, math, q-fin]}, February 2021.
\newblock arXiv: 1911.03380.

\bibitem{Aoyagi2020}
Jun Aoyagi.
\newblock Lazy {Liquidity} in {Automated} {Market} {Making}.
\newblock {\em SSRN Electronic Journal}, 2020.

\bibitem{bancorv2}
Bancor.
\newblock Proposing bancor v2.1: Single-sided amm with elastic bnt supply.
\newblock
  \url{https://blog.bancor.network/proposing-bancor-v2-1-single-sided-amm-with-elastic-bnt-supply-bcac9fe655b},
  October 2020.
\newblock As of 8 February 2021.

\bibitem{BartolettiCL2021}
Massimo Bartoletti, James Hsin-yu Chiang, and Alberto Lluch-Lafuente.
\newblock A theory of {Automated} {Market} {Makers} in {DeFi}.
\newblock {\em arXiv:2102.11350 [cs]}, April 2021.
\newblock arXiv: 2102.11350.

\bibitem{bichuch2023axioms}
Maxim Bichuch and Zachary Feinstein.
\newblock Axioms for automated market makers: A mathematical framework in
  fintech and decentralized finance, 2023.

\bibitem{capponi2021adoption}
Agostino Capponi and Ruizhe Jia.
\newblock The adoption of blockchain-based decentralized exchanges.
\newblock {\em arXiv preprint arXiv:2103.08842}, 2021.

\bibitem{cvetkovski2012inequalities}
Zdravko Cvetkovski.
\newblock {\em Inequalities: theorems, techniques and selected problems}.
\newblock Springer Science \& Business Media, 2012.

\bibitem{curve}
Michael Egorov.
\newblock Stableswap - efficient mechanism for stablecoin liquidity.
\newblock \url{https://www.curve.fi/stableswap-paper.pdf}, November 2019.
\newblock As of 8 Februaary 2021.

\bibitem{EngelH2021}
Daniel Engel and Maurice Herlihy.
\newblock Composing {Networks} of {Automated} {Market} {Makers}.
\newblock {\em ArXiv}, June 2021.
\newblock arXiv: 2106.00083.

\bibitem{uniswapv3}
{Hayden Adams}, {Noah Zinsmeister}, {Moody Salem}, {River Keefer}, and {Dan
  Robinson}.
\newblock Uniswap v3 {Core}, March 2021.

\bibitem{bancor}
Eyal Hertzog, Guy Benartzi, and Galia Benartzi.
\newblock Bancor protocol, 2017.

\bibitem{balancer}
Fernando Martinelli and Nikolai Mushegian.
\newblock Balancer: A non-custodial portfolio man- ager, liquidity provider,
  and price sensor.
\newblock https://balancer.finance/whitepaper/, 2109.
\newblock As of 2 February 2021.

\bibitem{MilionisMRZ2022}
Jason Milionis, Ciamac~C. Moallemi, Tim Roughgarden, and Anthony~Lee Zhang.
\newblock Automated {Market} {Making} and {Loss}-{Versus}-{Rebalancing},
  September 2022.
\newblock arXiv:2208.06046 [math, q-fin].

\bibitem{RamseyerGGM2023}
Geoffrey Ramseyer, Mohak Goyal, Ashish Goel, and David Mazières.
\newblock Augmenting batch exchanges with constant function market makers,
  2023.

\bibitem{xu2023sok}
Jiahua Xu, Krzysztof Paruch, Simon Cousaert, and Yebo Feng.
\newblock Sok: Decentralized exchanges (dex) with automated market maker (amm)
  protocols.
\newblock {\em ACM Computing Surveys}, 55(11):1--50, 2023.

\bibitem{zhang2018}
Yi~Zhang, Xiaohong Chen, and Daejun Park.
\newblock Formal specification of constant product (x . y = k) market maker
  model and implementation.
\newblock
  \url{https://github.com/runtimeverification/verified-smart-contracts/blob/uniswap/uniswap/x-y-k.pdf},
  2018.

\end{thebibliography}

\appendix
\section{Aggregation Mechanics}
Here, we will show that the weighted geometric mean aggregation algorithm described by equation \ref{eqn:c} is the unique valid aggregation algorithm. To this end, fix an arbitrary finite nonempty set of LPs $L$, and let 
\[
\mathtt{aggregate_0}((c_\ell)_{\ell \in L}, (s_\ell)_{\ell \in L}) = \prod_{\ell \in L}c_\ell^{s_\ell}.
\]
Recall the definition of a \emph{valid} aggregation algorithm, which we rewrite here for convenience.
\begin{definition}\label{aggregation_appendix}
    Let $\mathtt{aggregate}()$ be an aggregation algorithm; let $L$ be the particular set of active LPs at some time, and consider the corresponding aggregation function. We say that $\mathtt{aggregate}()$ is \emph{valid} if the following axioms hold.
\begin{enumerate}
    \item For each $\ell \in L$, $\mathtt{aggregate}$ has continuous partial derivatives with respect to $c_\ell$. Additionally, 
    \[
    \frac{\partial (\mathtt{aggregate})}{\partial\log(c_\ell)} = s_\ell \cdot \mathtt{aggregate}((c_\ell)_{\ell \in L}, (s_\ell)_{\ell \in L}).
    \]
    \item 
    If there exists some $x$ such that $c_\ell=x$ for all $\ell \in L$, then 
    \[
    \mathtt{aggregate}((c_\ell)_{\ell \in L}, (s_\ell)_{\ell \in L}) = x.
    \]
\end{enumerate}
\end{definition}

\begin{proposition}
    $\mathtt{aggregate_0}()$ defines a valid aggregation algorithm.
\end{proposition}
\begin{proof}
    Fix some $\ell \in L$. We write $c = \mathtt{aggregate_0}((c_k)_{k \in L}, (s_k)_{k \in L})$. Observe that
    \begin{align*}
    \frac{\partial(\mathtt{aggregate_0})}{\partial \log c_\ell} 
    &= \frac{\frac{\partial (\mathtt{aggregate_0})}{\partial c_\ell}}{\frac{\partial \log c_\ell}{\partial c_\ell}}\\
    &= \frac{\frac{\partial}{\partial c_\ell} \prod_{k \in L}c_k^{s_k}}{\frac{1}{c_\ell}}\\
    &= c_\ell s_\ell \cdot c_\ell^{s_\ell-1} \cdot \prod_{k \in L \setminus \{\ell\}} c_k^{s_k}\\
    &= s_\ell \prod_{k \in L} c_k^{s_k} = s_\ell c.
    \end{align*}
    This shows condition 1 of Definition \ref{aggregation_appendix}. Finally, consider any $((c_k)_{k \in L}, (s_k)_{k \in L})$ in the domain of $\mathtt{aggregate_0}$ such that there exists some $x$ such that $c_\ell = x$ for all $\ell \in L$. Since $\sum_{k \in L} s_k = 1$,
    \[
    c = \prod_{k \in L} c_k^{s_k} = \prod_{k \in L} x^{s_k} = x,
    \]
    as desired.
\end{proof}

Now we prove the uniqueness of $\mathtt{aggregate_0}()$.
\begin{proposition}
    $\mathtt{aggregate_0}()$ is the only valid aggregation algorithm.
\end{proposition}
\begin{proof}
    Let $\mathtt{aggregate}()$ be any valid aggregation algorithm. It suffices to show that $\mathtt{aggregate} = \mathtt{aggregate_0}$ as functions, for all nonempty finite sets $L$. To this end, let $L$ be finite and nonempty. Consider the function of the algorithm $\mathtt{aggregate}()$ on LP set $L$, which we call $c$ for brevity. Fix $(s_\ell)_{\ell \in L} \in [0,1]^L$ such that $\sum_{\ell \in L} s_\ell = 1$. Since $\mathtt{aggregate}$ is valid, for all $\ell \in L$,
    \[
    \frac{\partial c}{\partial\log(c_\ell)} = s_\ell \cdot c.
    \]
    By the chain rule,
    \begin{align*}
        \frac{\partial c}{\partial \log c_\ell} 
        &= \frac{\frac{\partial c}{\partial \log c_\ell}}{\frac{\partial \log c_\ell}{\partial c_\ell}}\\
        &= c_\ell \cdot \frac{\partial c}{\partial c_\ell}.
    \end{align*}
    Thus 
    \begin{align*}
        \frac{\partial c}{\partial c_\ell} &= \frac{s_\ell}{c_\ell} \cdot c\\
        \implies \int \frac{dc}{c} &= \int s_\ell \cdot \frac{dc_\ell}{c_\ell}\\
        \implies \log(c) &= s_\ell \log(c_\ell) + A((c_k)_{k \in L \setminus \{\ell\}})\\
        &= \log(c_\ell^{s_\ell}) + A((c_k)_{k \in L \setminus \{\ell\}})\\
        \implies c &= c_\ell^{s_\ell} \cdot B((c_k)_{k \in L \setminus \{\ell\}})
    \end{align*}
    for some functions $A((c_k)_{k \in L \setminus \{\ell\}})$ and $B((c_k)_{k \in L \setminus \{\ell\}})$
    This implies that there exists a constant $C$ (possibly dependent on $(s_\ell)_{\ell \in L}$) such that
    \[
    c = C \cdot \prod_{\ell \in L} c_\ell^{s_\ell}.
    \]
    Since this holds for any $(c_\ell)_{\ell \in L}$, we may choose $(c_\ell)_{\ell \in L}$ so that $c_\ell = x$ for all $\ell \in L$, for a fixed $x > 0$.
    Using the second condition of Definition \ref{aggregation_appendix}, we know that 
    \[
    c((c_\ell)_{\ell \in L}, (s_\ell)_{\ell \in L}) = x.
    \]
    This implies
    \begin{align*}
        x &= c((c_\ell)_{\ell \in L}, (s_\ell)_{\ell \in L})\\
        &= C \cdot \prod_{\ell \in L} c_\ell^{s_\ell}\\
        &= C \cdot \prod_{\ell \in L} x^{s_\ell}\\
        &= C \cdot x
    \end{align*}
    Since $x > 0$, it follows that $C = 1$. Thus, for any $((c_\ell)_{\ell \in L}, (s_\ell)_{\ell \in L})$ in the domain of $c$, it follows that
    \[
    c((c_\ell)_{\ell \in L}, (s_\ell)_{\ell \in L}) = \prod_{\ell \in L} c_\ell^{s_\ell},
    \]
    as desired.
\end{proof}

We conclude by proving some basic properties of $\mathtt{aggregate_0}()$.
\begin{proposition}
    Fix a finite nonempty set $L$. Fix $(s_\ell)_{\ell \in L}$ satisfying $\sum_{\ell \in L} s_\ell = 1$. Then the corresponding function of $\mathtt{aggregate_0}()$, simply denoted $\mathtt{aggregate_0}$, satisfies the following.
    \begin{itemize}
        \item (Continuity.) $\mathtt{aggregate_0}$ is continuous in each coordinate.
        \item (Increasing.) $\mathtt{aggregate_0}$ is weakly increasing in each $c_\ell$, and strictly increasing in $c_\ell$ if $s_\ell > 0$.
        \item (Convexity.) 
        For all $((c_\ell)_{\ell \in L}, (s_\ell)_{\ell \in L})$ in the domain of $\mathtt{aggregate_0}$,
        \[
        \mathtt{aggregate_0}((c_\ell)_{\ell \in L}, (s_\ell)_{\ell \in L}) \in \conv \{c_\ell: \ell \in L\}.
        \]
        \item (Boundary conditions.) If $s_\ell = 1$, then $\mathtt{aggregate_0}((c_\ell)_{\ell \in L}, (s_\ell)_{\ell \in L}) = c_{\ell}$.
    \end{itemize}
\end{proposition}
\begin{proof}
    The Continuity, Increasing, and Boundary conditions properties are obvious. For the Convexity property, let $c = \mathtt{aggregate_0}((c_\ell)_{\ell \in L}, (s_\ell)_{\ell \in L})$.
    Let $c_{\mathtt{min}} = \min\{c_k: k\in L\}$ and $c_{\mathtt{max}} = \max\{c_k:k \in L\}$. Since $\sum_{k \in L} s_k = 1$, it follows that
    \[
    c_{\mathtt{min}} = \prod_{k \in L} c_{\mathtt{min}}^{s_k} \le \prod_{k \in L} c_k^{s_k} \le \prod_{k \in L} c_{\mathtt{max}}^{s_k} = c_{\mathtt{max}}.
    \]
    Since $c = \prod_{k \in L} c_k^{s_k}$, it follows that $c \in \mathrm{conv} \{c_k : k \in L\}$.
\end{proof}

\end{document}